\documentclass[prd,twocolumn,floatfix,amsmath,nofootinbib,amssymb,floatfix]{revtex4}
\usepackage{graphicx,color,dcolumn,booktabs,bm}
\usepackage{longtable,lscape}
\usepackage{txfonts}
\usepackage{overpic}
\usepackage{amssymb}
\usepackage{indentfirst}
\usepackage{feynmf}   
\usepackage{slashed}  
\usepackage{cases}
\usepackage{color}
\usepackage{multirow}
\usepackage{epstopdf}
\usepackage{enumerate}
\usepackage{graphicx,color,dcolumn,booktabs,bm}
\usepackage[normalem]{ulem}
\usepackage[colorlinks, citecolor=blue,anchorcolor=red,menucolor=red, linkcolor=red,filecolor=red,urlcolor=blue,frenchlinks=red]{hyperref}

\graphicspath{{Figures/}} %

\begin{document}

\title{Evidence supporting the existence of $P_c(4380)^{\pm}$ from the recent measurements of $B_s \to J/\psi p\bar{p}$}
\author{Jun-Zhang Wang$^{1,2}$}\email{wangjzh2012@lzu.edu.cn}
\author{Xiang Liu$^{1,2,3}$\footnote{Corresponding author}}\email{xiangliu@lzu.edu.cn}
\author{Takayuki Matsuki$^{4}$}\email{matsuki@tokyo-kasei.ac.jp}
\affiliation{$^1$School of Physical Science and Technology, Lanzhou University, Lanzhou 730000, China\\
$^2$Research Center for Hadron and CSR Physics, Lanzhou University $\&$ Institute of Modern Physics of CAS, Lanzhou 730000, China\\
$^3$Lanzhou Center for Theoretical Physics, Key Laboratory of Theoretical Physics of Gansu Province, and Frontier Science Center for Rare Isotopes, Lanzhou University, Lanzhou 730000, China\\ 
$^4$Tokyo Kasei University, 1-18-1 Kaga, Itabashi, Tokyo 173-8602, Japan}

\date{\today}

\begin{abstract}
Very recently, the LHCb collaboration released the newest measurements of $B_s \to J/\psi p\bar{p}$, where an enhancement structure near 4.34 GeV was observed in the invariant mass spectrum of $J/\psi p$ with the statistical significance of $3.1-3.7$ $\sigma$. In this work, by performing a combined analysis for the three invariant mass spectra of $B_s \to J/\psi p\bar{p}$, we find that this $J/\psi p$ structure near 4.34 GeV can correspond to the contributions from the $P_c(4380)^+$ state with the assumption of $J^{P}=3/2^{-}$ together with the reflections of $B_s \to (f_2^{*} \to p\bar{p})J/\psi$. Here, $P_c(4380)$ was first observed in $\Lambda_b \to J/\psi p K$ but not confirmed in the updated LHCb data in 2019, and $f_2^{*}$ means the exciting light isoscalar tensor meson around 2.0 GeV. Thus, this provides a possible evidence to support the existence of the pentaquark $P_c(4380)$. Specifically, the assumed spin parity $J^{P}=3/2^{-}$ of $P_c(4380)$ is consistent with a prediction of the theoretical explanation of an $S$-wave $\bar{D}\Sigma_c^*$ molecular bound state.
\end{abstract}

\maketitle


\section{Introduction}

Very recently, the LHCb collaboration announced the observation of a new pentaquark-like structure $P_c(4337)^+$ with a relatively low statistical significance of $3.1-3.7$ $ \sigma$ in the $J/\psi p$ invariant mass spectrum of $B_s \to J/\psi p\bar{p}$ \cite{LHCb:2021chn}. The resonance parameters of this newly observed $P_c$ structure are different from those of the currently known pentaquark states reported in the  $\Lambda_b \to J/\psi p K$ by LHCb, which include  $P_c(4312)^+$, $P_c(4440)^+$, and $P_c(4457)^+$ \cite{LHCb:2019kea}. Specifically, these three $P_c$ structures that are all slightly below the thresholds of di-hadrons composed of an S-wave charmed baryon $\Sigma_c$  and an S-wave anticharmed meson $\bar{D}$ or $\bar{D}^*$ have provided strong evidence for the existence of the hidden-charm molecular pentaquarks, where $P_c(4312)^+$, $P_c(4440)^+$, and $P_c(4457)^+$ can just correspond to the $S$-wave loosely bound states of the $\bar{D}\Sigma_c$ with $J^P=1/2^{-}$, $\bar{D}^*\Sigma_c$ with $J^P=1/2^{-}$, and $J^P=3/2^{-}$ \cite{Wu:2010jy,Wang:2011rga,Yang:2011wz,Uchino:2015uha,Karliner:2015ina,Chen:2019asm,Liu:2019tjn,Yamaguchi:2019seo,Chen:2019bip,Xiao:2019aya,Meng:2019ilv,PavonValderrama:2019nbk,He:2019ify,Du:2019pij,Burns:2019iih,Wang:2019ato}, respectively. Therefore, it seems that this newly observed $P_c(4337)$ cannot be assigned to the same theoretical configuration with the above three $P_c$ states. In this situation, it is a challenging task for theorists to understand the origin of $P_c(4337)$.
 
After the release of measurements of $B_s \to J/\psi p\bar{p}$, there have been several theoretical works to discuss the nature of $P_c(4337)$ \cite{Yan:2021nio,Liu:2021ixf,Liu:2021efc,Nakamura:2021dix,Giron}. In Ref. \cite{Yan:2021nio}, Yan $et$ $al.$ proposed three possible theoretical explanations, i.e., a hadrocharmonium of $\chi_{c0}p$ bound state, $\bar{D}^*\Lambda_c$ and $\bar{D} \Sigma_c$ state close to threshold and $\bar{D}^*\Lambda_c$ and $\bar{D} \Sigma_c^*$ coupled channel dynamics. Although the magnitude of $P_c(4337)$ mass might be explained in these configurations, there appears another problem that it is not clear why this structure is not observed in  decay process $\Lambda_b \to J/\psi p K$ with large reconstructed events \cite{LHCb:2019kea}. The authors in Ref. \cite{Nakamura:2021dix} suggested that the $P_c(4312)$  and  $P_c(4337)$ can be created by different interference patterns between the $\Sigma_c\bar{D}$ and $\Lambda_c\bar{D}^*$ threshold cusps, which can give a reasonable interpretation to the phenomenon that the $P_c(4312)$ and $P_c(4337)$ peaks appear in $\Lambda_b \to J/\psi p K$ and $B_s \to J/\psi p\bar{p}$, respectively.

In this work, we propose a completely different perspective to decode the nature of $P_c(4337)$.
Here is an interesting possibility that the peak position of this $J/\psi p$ structure near 4.34 GeV may deviate from its genuine mass because this structure is close to the upper threshold of the $J/\psi p$ phase space in $B_s \to J/\psi p\bar{p}$. In fact, there exists a potential candidate to explain the LHCb observation without introducing a new pentaquark state $P_c(4337)$. Let us recall $P_c(4380)$, which was first reported in the $J/\psi p$ mass spectrum of $\Lambda_b \to J/\psi p K$ in 2015  \cite{LHCb:2015yax}. Unfortunately, this state cannot be confirmed in the 2019 updated data in the same decay process \cite{LHCb:2019kea}, so it will be very interesting and meaningful to explore this possibility that there could be influence of $P_c(4380)$ on the LHCb result. Of course, another advantage of this view is that it can provide a natural interpretation to connect the observations in the $J/\psi p$ mass distribution of $B_s \to J/\psi p\bar{p}$ and $\Lambda_b \to J/\psi p K$.  
In addition to this possibility, we also notice that there are several suspected signals in the $p\bar{p}$ invariant mass spectrum of  $B_s \to J/\psi p\bar{p}$ despite the low statistical significance \cite{LHCb:2021chn}. Interestingly, the contribution of the intermediate light meson decays into $p\bar{p}$ may be reflected on the $J/\psi p$ and $J/\psi \bar{p}$ mass spectrum as the line shape of the reflection peaks, which can usually mimic the resonance signals. This reflection mechanism has been successfully applied in explaining some charmoniumlike $XYZ$ states \cite{Wang:2020axi,Wang:2020dmv,Wang:2020kej}. Hence, what role do the reflection contributions of possible light mesonic states play in depicting this new $J/\psi p$ structure near 4.34 GeV is also worth investigating.

By a combined fit to the LHCb data of three invariant mass spectra of $B_s \to J/\psi p\bar{p}$, we will show later that the $P_c(4337)$ structure can indeed be reproduced by including the contributions of $P_c(4380)$ with the assumption of $J^{P}=3/2^{-}$ together with the reflections from the exciting light isoscalar tensor mesons around 2.0 GeV. Furthermore, we find that our predictions for the angular distributions of $B_s \to J/\psi p\bar{p}$ can also meet the LHCb data well. Therefore, this means that our perspective shows a possible new evidence to support the existence of the $P_c(4380)$ state from the measurement of $B_s \to J/\psi p\bar{p}$, which should provide some valuable hints for constructing the exotic pentaquark hadron family.

This paper is organized as follows. After the Introduction, we study the reflection line shapes from the intermediate light scalar and tensor mesons on the $J/\psi p$ mass spectrum of $B_s \to J/\psi p\bar{p}$ and the corresponding $J/\psi p$ mass distribution of the $P_c(4380)$ contribution with two assumptions of $J^{P}=1/2^{-}$ and $3/2^{-}$  in Sec. \ref{sec2}. According to the research findings in Sec. \ref{sec2}, we perform a combined fit to three invariant mass spectra of $B_s \to J/\psi p\bar{p}$ in Sec. \ref{sec3}, where we find that the $P_c(4380)$ state with $J^{P}=3/2^{-}$ can indeed play an important role in describing the line shape of the $J/\psi p$ enhancement around 4.34 GeV. This paper ends with a conclusion in Sec. \ref{sec4}.

\section{The light meson reflections and the role of pentaquark $P_c(4380)$ in $B_s \to J/\psi p\bar{p}$ }\label{sec2}

The $B/B_s$ meson decay is a very important experimental platform to search for new hadronic structures. Since 2003, many famous charmoniumlike structures were discovered in the $B$ meson decay processes with three-body final states, such as $X(3872)$ \cite{Belle:2003nnu}, $Y(3940)$ \cite{Belle:2004lle},  $Z(4430)^+$ \cite{Belle:2007hrb}, $Y(4140)$ \cite{CMS:2013jru} as well as seven $X$ structures in the $J/\psi \phi$ spectrum  and  $Z_{cs}(4000)^+$ and $Z_{cs}(4220)^+$ in the $J/\psi K^+$ spectrum recently reported  in the $B^+ \to J/\psi \phi K^+$ \cite{LHCb:2021uow}. These surprising observations have attracted a lot of attention from theorists and experimentalists. However, it is still a very great challenge for theorists to explain such rich exotic structures seen in the $B$ meson decays in a unified theoretical framework \cite{Wang:2021aql,Chen:2016qju,Liu:2019zoy}.  Here, an available way to alleviate the present situation is the so-called nonresonant explanation \cite{Chen:2016qju}.


As one of the nonresonant explanations, the reflection mechanism in a general decay cascade process of $ H\to AB \to A (B \to CD)$, where $H$ is an initial state particle, is that the distribution of an intermediate resonance $B$ on the invariant mass spectrum of $CD$ can be reflected into the other mass spectra of $AC$ and $AD$ as the line shapes of reflection peaks when the system satisfies some special kinematic and dynamical conditions \cite{Wang:2020dmv}, which may mimic the  signal of the announced exotic hadronic resonances. The reflection mechanism has been succeeded in explaining several charmoniumlike structures observed in the electron-positron annihilation, such as $Z_c(3885)$ \cite{Wang:2020axi}, $Z_c(4025)$ \cite{Wang:2020axi}, and $Z_{cs}(3985)$ \cite{Wang:2020kej}.

If carefully checking the LHCb data of $B_s \to J/\psi p\bar{p}$ \cite{LHCb:2021chn}, one can actually see several suspected enhancements or dips near 2.0 and 2.2 GeV in the $p\bar{p}$ mass spectrum, which are ignored in the experimental analysis because of the low statistics of the data. Thus, these interesting clues inspire us to explore whether the reflections from higher light meson states play the important roles in producing the  $P_c(4337)$ structure seen in the mass spectrum of $J/\psi p$. By consulting the Particle Data Group (PDG) \cite{ParticleDataGroup:2020ssz},  we find that the productions of the ground states and many excited states of light isoscalar scalar and tensor mesons accompanied by a $J/\psi$ in the $B_s$ meson decay have been confirmed such as $B_s \to (f_0(980)/f_0(1790) \to \pi^+\pi^-) J/\psi$ \cite{Belle:2011phz,LHCb:2014ooi} and $B_s \to (f_2(1270)/f_2^{\prime}(1525) \to \pi^+\pi^-) J/\psi$ \cite{LHCb:2012ae,LHCb:2014ooi}.  Hence, we mainly consider the scalar and tensor light mesons with quantum numbers $J^{P}=0^+$ and $2^+$ in this work, respectively.
Of course, in addition to the reflection contributions from possible highly excited light meson decays into $p\bar{p}$, there also exist two other production mechanisms, i.e.,  $B_s \to (P_c^- \to J/\psi \bar{p})  p$ and  $B_s \to (P_c^+ \to J/\psi p) \bar{p}$, where $P_c$ is an exotic pentaquark state. Before introducing a new pentaquark, we conjecture that this $J/\psi p$ enhancement around 4.34 GeV can also be caused by a nearby pentaquark state because the resonance that is close to the threshold may deviate from the standard Breit-Wigner distribution.  Here, the narrow $P_c(4312)$ state, which was first established in the 2019 updated measurement of $\Lambda_b \to J/\psi p K$ \cite{LHCb:2019kea}, is near the 4.34 GeV. In addition, there exists another candidate of a pentaquark reported in the $J/\psi p$ invariant mass spectrum, which is named the $P_c(4380)$  in the measurement of the same decay process of $\Lambda_b \to J/\psi p K$ in 2015 \cite{LHCb:2015yax}. However, there is a serious problem that this more broad state was not further confirmed in the 2019 LHCb measurement \cite{LHCb:2019kea}. Therefore, it will be an intriguing topic to explore the role of $P_c(4380)$ in describing the line shape of the $J/\psi p$ mass distribution in $B_s \to J/\psi p\bar{p}$.

\begin{figure}[t]
	\includegraphics[width=8.8cm,keepaspectratio]{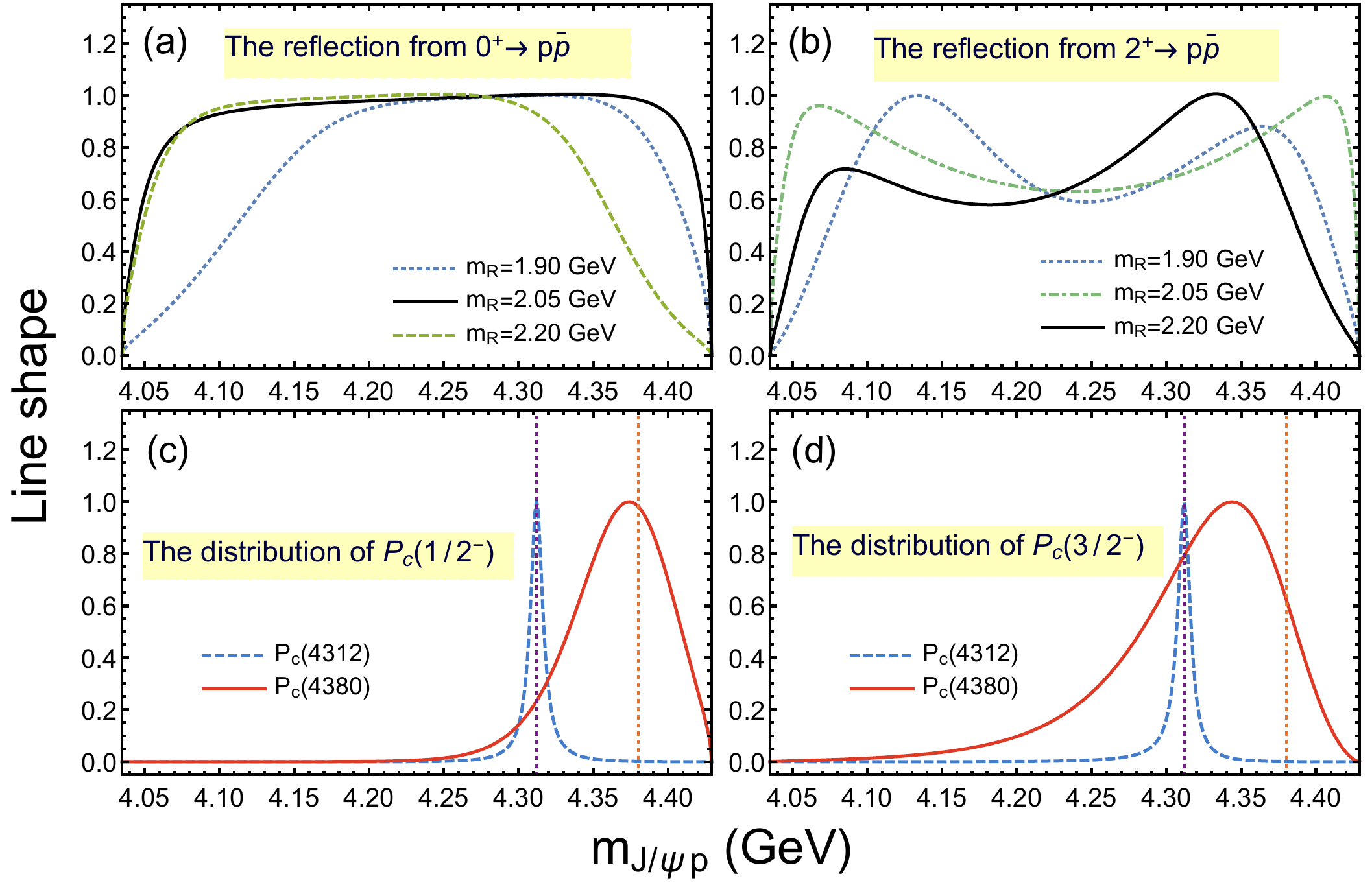}
	\caption{ The $J/\psi p$ mass distributions of $B_s \to J/\psi p\bar{p}$ from the reflection contributions of scalar and tensor states with three typical mass values of $m_R=1.90, 2.05$, and 2.20 GeV as well as the pentaquark contributions  of $P_c(4312)$ and $P_c(4380)$ with the assumption of $J^{P}=1/2^{-}$ and $3/2^{-}$.  Here, the dotted lines in (c) and (d) mean the respective mass positions of $P_c$ resonances. \label{fig:jpsipbarspectrum} }
\end{figure}

For studying these underlying contributions in $B_s \to J/\psi p\bar{p}$ as mentioned above, we adopt  the effective Lagrangian approach.   The relevant Lagrangian densities  are listed below \cite{Tsushima:1996xc,Tsushima:1998jz,Zou:2002yy,Wu:2009md,Wang:2017sxq,Colangelo:2010te,Yan:1999fn,Goldberg:1968zza}:
\begin{eqnarray}
\mathcal{L}_{\mathcal{R}_0NN}&=&g_{\mathcal{R}_0NN}\bar{N}\mathcal{R}_0N,    \\
\mathcal{L}_{\mathcal{R}_2NN}&=&g_{\mathcal{R}_2NN}\partial^{\mu}\bar{N}\partial^{\nu}N\mathcal{R}_{2\mu\nu},    \\
\mathcal{L}_{\psi NP_c}&=&-g_{\psi NP_c}(\bar{N}\gamma_{\mu}\psi^{\mu}P_c+\bar{P}_c\gamma_{\mu}\psi^{\mu}N), \\
\mathcal{L}_{\psi NP_c^{*}}&=&-g_{\psi NP_c^{*}}(\bar{N}\psi_{\mu}P_{c}^{*\mu}+\bar{P}_{c}^{*\mu}\psi_{\mu}N),  \\
\mathcal{L}_{B_s\mathcal{R}_0\psi}&=&g_{B_s\mathcal{R}_0\psi}B_s\partial_{\mu}\mathcal{R}_0\psi^{\mu},    \\
\mathcal{L}_{B_s\mathcal{R}_2\psi}&=&g_{B_s\mathcal{R}_2\psi}\partial_{\mu}B_s\partial_{\lambda}\mathcal{R}_2^{\mu\nu}\partial_{\nu}\psi^{\lambda},    
\end{eqnarray}
\begin{eqnarray}
\mathcal{L}_{B_s NP_c}&=&-g_{B_s NP_c}(\bar{N}\gamma_5\gamma^{\mu}P_c\partial_{\mu}B_s-\bar{P}_c\gamma_5\gamma^{\mu}N\partial_{\mu}B_s), \\
\mathcal{L}_{B_s NP_c^{*}}&=&-g_{B_s NP_c^{*}}(\bar{N}\gamma_5\gamma^{\mu}P_c^{*\nu}\partial_{\mu}\partial_{\nu}B_s+\bar{P}_c^{*\nu}\gamma_5\gamma^{\mu}N\partial_{\mu}\partial_{\nu}B_s),  \nonumber \\
\end{eqnarray}
where $\mathcal{R}_0$ and $\mathcal{R}_2$ are the light meson fields with $J^{P}=0^+$ and $2^+$, respectively, and $P_c$ and $P_c^{*\nu}$ stand for the pentaquark fields with $J^{P}=1/2^-$ and $3/2^-$, respectively.  By the above Lagrangian densities,  the decay amplitudes of the  $B_s(p_1) \to \mathcal{R}_{0/2}(q)J/\psi(p_3) \to p(p_4)\bar{p}(p_5)J/\psi(p_3)$ can be written as
\begin{eqnarray}
\mathcal{A}(B_s \to (\mathcal{R}_{0}\to p\bar{p})J/\psi)&=&\frac{g_{1}(q\cdot \epsilon_{\psi} )\bar{u}(p_4)v(p_5)\mathcal{F}(q^2,m_{\mathcal{R}_{0}})}{q^2-m^2_{\mathcal{R}_{0}}+im_{\mathcal{R}_{0}}\Gamma_{\mathcal{R}_{0}}}, \nonumber \\
\mathcal{A}(B_s \to (\mathcal{R}_{2}\to p\bar{p})J/\psi)&=&\frac{g_{2}(q\cdot \epsilon_{\psi})\bar{u}(p_4)v(p_5)\mathcal{F}(q^2,m_{\mathcal{R}_{2}}^2)}{q^2-m^2_{\mathcal{R}_{2}}+im_{\mathcal{R}_{2}}\Gamma_{\mathcal{R}_{2}}}  \nonumber \\
&&\times G_{\mu\alpha\nu\beta}p_4^{\mu}p_5^{\alpha}p_1^{\nu}p_3^{\beta},
\end{eqnarray}
where $G_{\mu\alpha\nu\beta}=\frac{1}{2}(\tilde{g}_{\mu\nu}\tilde{g}_{\alpha\beta}+\tilde{g}_{\mu\beta}\tilde{g}_{\alpha\nu})-\frac{1}{3}\tilde{g}_{\mu\alpha}\tilde{g}_{\nu\beta}$ with $\tilde{g}_{\mu\nu}=-g_{\mu\nu}+q_{\mu}q_{\nu}/m_{\mathcal{R}_{2}}^2$ is the spin projection operator of tensor particle, and $g_{1}/g_{2}$ absorbs all constant terms in the amplitude.  Here, the form factor $\mathcal{F}(q^2,m_{\mathcal{R}})=\Lambda^4/(\Lambda^4+(q^2-m_{\mathcal{R}}^2)^2)$ is introduced to describe the off-shell effect of the intermediate resonance state \cite{Feuster:1997pq,Haberzettl:1998aqi,Yoshimoto:1999dr,Oh:2000zi}, and the cutoff $\Lambda$ will be taken as one in the following calculations. Additionally, the decay amplitudes of the  $B_s(p_1) \to P_c^{(*)+}(q^{\prime})\bar{p}(p_5) \to J/\psi(p_3)p(p_4)\bar{p}(p_5)$ can be given by
\begin{eqnarray}
\mathcal{A}(B_s \to (P_c^{+}\to J/\psi p)\bar{p})&=&\frac{g_{3}\bar{u}(p_4)\gamma_{\nu}({q}\!\!\!{\slash}^{\prime}+m_{P_c})\gamma_5\gamma_{\mu}v(p_5)}{q^{\prime2}-m^2_{P_c}+im_{P_c}\Gamma_{P_c}} \nonumber \\
&&\times p_1^{\mu}\epsilon_{\psi}^{\nu}\mathcal{F}(q^{\prime2},m_{P_c}), \nonumber \\
\mathcal{A}(B_s \to (P_c^{*+}\to J/\psi p)\bar{p})&=&\frac{g_{4}\bar{u}(p_4)({q}\!\!\!{\slash}^{\prime}+m_{P_c^{*}})\gamma_5\gamma_{\lambda}v(p_5)}{q^{\prime2}-m^2_{P_c^{*}}+im_{P_c^{*}}\Gamma_{P_c^{*}}} \nonumber \\
&&\times \mathcal{E}_{\mu\nu}p_1^{\lambda}p_1^{\nu}\epsilon_{\psi}^{\mu}\mathcal{F}(q^{\prime2},m_{P_c^{*}}), \label{eqs-10}
\end{eqnarray}
where $\mathcal{E}_{\mu\nu}=-g_{\mu\nu}+\frac{1}{3}\gamma_{\mu}\gamma_{\nu}+\frac{1}{3m_{P_c^{*}}}(\gamma_{\mu}q^{\prime}_{\nu}-\gamma_{\nu}q^{\prime}_{\mu})+\frac{2q^{\prime}_{\mu}q^{\prime}_{\nu}}{3m_{P_c^{*}}^2}$. Similarly, the decay amplitudes of the  $B_s(p_1) \to P_c^{(*)-}(q^{\prime\prime})p(p_4) \to J/\psi(p_3)\bar{p}(p_5)p(p_4)$ can be directly obtained by replacing $q^{\prime}$ in Eq. (\ref{eqs-10})  with  $(p_1-p_4)$.
Then, three invariant mass distributions of the above decay processes can be expressed as
\begin{eqnarray}
\frac{d\sigma}{dm_{p\bar{p}}}=\frac{\left|\bf{p_4}\right|\left|\bf {p_5}^*\right|\left|\mathcal{A}\right|^2}{(2\pi)^516m_{B_s}^2}d\Omega_{4}d\Omega_{5}^* \label{eq:11}, \\
\frac{d\sigma}{dm_{J/\psi p}}=\frac{\left|\bf{p_3}\right|\left|\bf {p_4}^*\right|\left|\mathcal{A}\right|^2}{(2\pi)^516m_{B_s}^2}d\Omega_{3}d\Omega_{4}^* \label{eq:12}, \\
\frac{d\sigma}{dm_{J/\psi \bar{p}}}=\frac{\left|\bf{p_3}\right|\left|\bf {p_5}^*\right|\left|\mathcal{A}\right|^2}{(2\pi)^516m_{B_s}^2}d\Omega_{3}d\Omega_{5}^* \label{eq:13} ,
\end{eqnarray}
where $\bf {p^*_i}$ and $\Omega_{i}^*$ are the corresponding three-momentum and solid angle in the center of mass frame of the  invariant mass system, respectively.

The line shapes of differential decay widths of the $B_s \to J/\psi (\mathcal{R}_{0}/\mathcal{R}_{2}) \to J/\psi (p\bar{p})$ on the invariant mass $J/\psi p$  with three typical resonance parameters of $(m_R,\Gamma_R)=(1.90,0.1), (2.05,0.1)$ and (2.20,0.1) GeV are presented in Figs. \ref{fig:jpsipbarspectrum}(a) and \ref{fig:jpsipbarspectrum}(b). It can be seen that the reflections from light scalar mesons basically play a background role and light  tensor mesons can produce the line shape of a reflection peak  in the invariant mass distribution of $J/\psi p$, whose peak position depends on the resonance mass of an intermediate mesonic state.  Furthermore, we investigate the pentaquark contributions from narrow  $P_c(4312)$ and broad $P_c(4380)$ states with the assumption of $J^{P}=1/2^{-}$ and $3/2^{-}$ for the $J/\psi p$ invariant mass spectrum, which are shown in Figs. \ref{fig:jpsipbarspectrum}(c) and \ref{fig:jpsipbarspectrum}(d).  Interestingly, we find that the contribution of a narrow $P_c(4312)$ state shows a standard Breit-Wigner line shape, but the peak position of $P_c(4380)$ obviously deviates from its resonance mass. The reason for this is that  the $P_c(4380)$ state  is close to the upper limit of $J/\psi p$  phase space in $B_s \to J/\psi p\bar{p}$ and simultaneously has a relatively large width, whose invariant mass distribution could be more easily affected by the phase space function and amplitude structure. Thus, the more broad $P_c(4380)$ state  has the potential to explain the newly reported $J/\psi p$ structure around 4.34 GeV in $B_s \to J/\psi p\bar{p}$, and the possibility of $P_c(4312)$ can be basically excluded.

From Fig. \ref{fig:jpsipbarspectrum}(d), one can see that there is a significantly large discrepancy between a peak position and resonance mass  when assuming the quantum number $J^{P}$ of  $P_c(4380)$ to be $3/2^{-}$ compared with the case of $P_c(4380)$ with $J^{P}=1/2^{-}$. Specifically, the corresponding peak position of $P_c(4380)$ is found at 4340 MeV if taking the LHCb central value of $m_{P_c(4380)}=4380$ MeV and a  $\Gamma_{P_c(4380)}=205$ MeV \cite{LHCb:2015yax} as input.  It is worth noticing that this result corresponds exactly to the reported resonance mass of the $P_c(4337)^{\pm}$ structure  in decay process $B_s \to J/\psi p\bar{p}$ \cite{LHCb:2021chn}. Hence, this finding gives us enough confidence to speculate that this newly reported  $J/\psi p$ structure near 4.34 GeV  may be not a new pentaquark candidate, where the missing pentaquark $P_c(4380)$ in the 2019 updated data of $\Lambda_b \to J/\psi p K$ \cite{LHCb:2019kea} plays a considerable role. Of course, the premise of this argument is based on that the spin parity of $P_c(4380)$ should be $3/2^{-}$ instead of $1/2^{-}$.  In the next section, we will explore this possibility carefully by fitting the experimental data of $B_s \to J/\psi p\bar{p}$.

\begin{figure*}[t]
	\includegraphics[width=14cm,keepaspectratio]{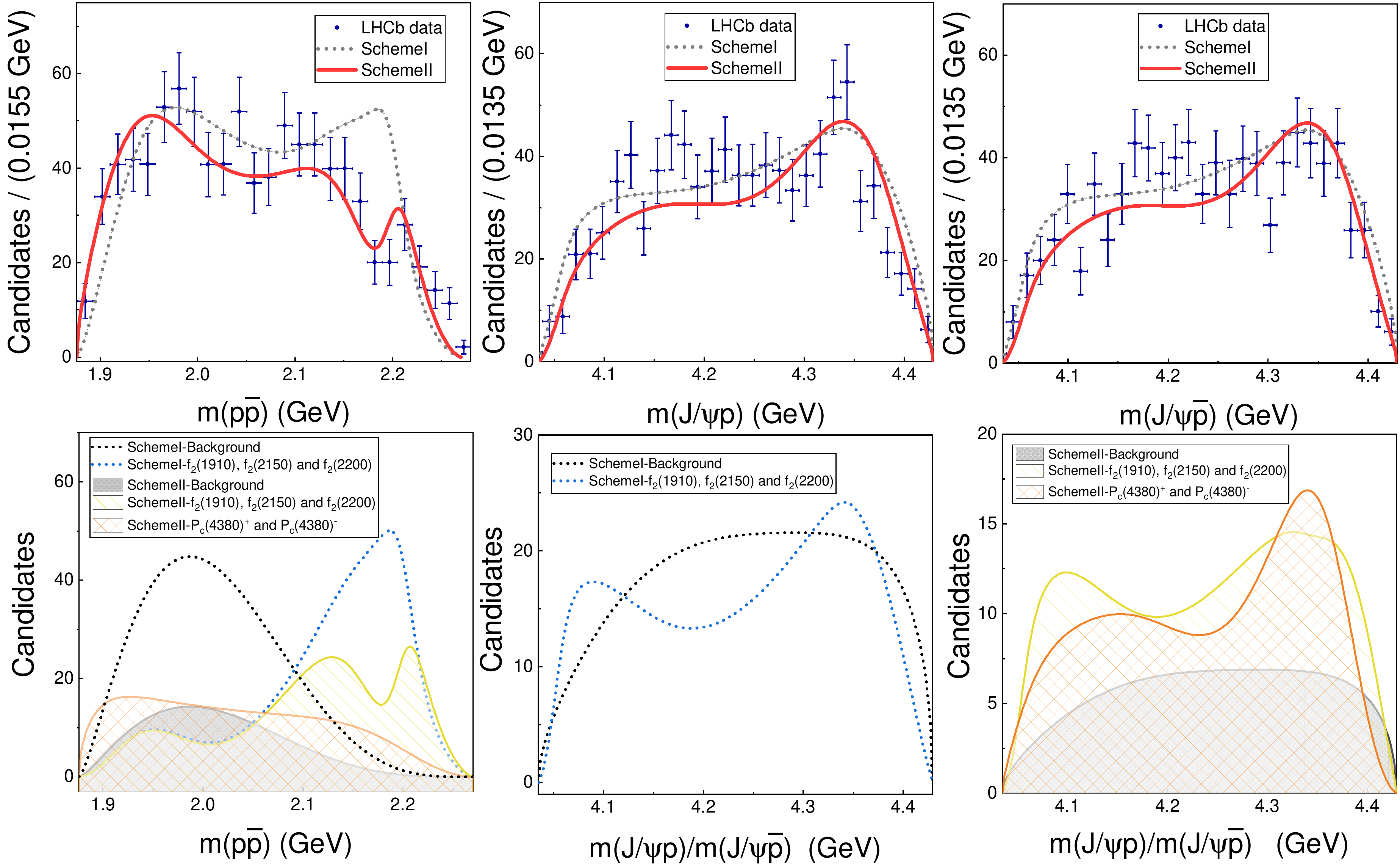}
	\caption{ The combined fit to the LHCb measurement of the invariant mass spectra of $p\bar{p}$, $J/\psi p$ and $J/\psi \bar{p}$ in the $B_s \to J/\psi p\bar{p}$. Here, the scheme-I and scheme-II fits are adopted, which correspond to the cases without and with the pentaquark contribution of $P_c(4380)^{\pm}$ with the assumption of $J^{P}=3/2^{-}$, respectively. \label{fig:fit} }
\end{figure*}

\section{Depicting the $P_c(4337)$  structure in the invariant mass spectrum of $J/\psi p$}\label{sec3}

As presented in the former section, the reflections from light tensor mesons or the $P_c(4380)$ state with $J^{P}=3/2^{-}$ can produce the peak line shape on the $J/\psi p$ mass spectrum,  both of which may explain the signal of $P_c(4337)$. In order to demonstrate the existence of $P_c(4380)$ in $B_s \to J/\psi p\bar{p}$, we first study the possibility to explain the $P_c(4337)$ signal by considering the reflection mechanism from  light-flavor $f_2$ mesonic states alone. Because the phase space of the $p\bar{p}$ mass spectrum in $B_s \to J/\psi p\bar{p}$ is limited to the range of $1876-2270$ MeV, we only need to investigate the $f_2$ meson states around 2.0 GeV.  In Ref. \cite{Ye:2012gu}, the Lanzhou Group had made a very comprehensive research on the $f_2$ meson family around 2.0 GeV, whose concise information is summarized in Table \ref{table:f2}. It is worth emphasizing that $f_2(2200)$ is a predicted radially excited state from the Regge tragectory extrapolation, whose width is estimated to be about $30-70$ MeV when $R=4.0-4.4$ GeV$^{-1}$ \cite{Ye:2012gu}. Here, we adopted a typical resonance parameter of $m=2.2$ GeV and $\Gamma=0.05$ GeV for $f_2(2200)$. As a matter of fact, the total widths of light isoscalar tensor mesons near 2.0 GeV are generally smaller than those of other light meson states such as light isoscalar scalar mesons \cite{ParticleDataGroup:2020ssz}. Thus, it is not easy to distinguish the contributions of other light mesons from the direct background. Based on this point, in the following analysis, we mainly focus on the reflections from three isoscalar tensor states $f_2(1910)$, $f_2(2150)$ and $f_2(2200)$.  For the reason of  decreasing the fitting parameters, we do not consider the $f_2(2010)$ state, which is a candidate of an $F$-wave state and this high partial wave contribution is assumed to be suppressed. Additionally, we also notice that these $f_2$ states have been experimentally observed in low energy $p\bar{p}$ collisions \cite{CrystalBarrel:2006mhj,Uman:2006xb,CrystalBarrel:1999hmx,Dulude:1978kt,Evangelista:1997be}.  Thus, this indicates that it should be reasonably enough to attribute several suspected signals seen in the measured $p\bar{p}$ invariant mass spectrum of $B_s \to J/\psi p\bar{p}$ \cite{LHCb:2021chn} to the resonance contributions from the $f_2$ mesons around 2.0 GeV. In the concrete fit to the invariant mass spectra of $B_s \to J/\psi p\bar{p}$, we still need a direct coupling term in the total decay amplitude, which can be written as \begin{eqnarray}
\mathcal{A}^{\textrm{Total}}=\mathcal{A}^{\textrm{Nonpeak}}+\sum_i e^{i\phi_i}\mathcal{A}(B_s \to (f_{2}^{(i)}\to p\bar{p})J/\psi) \label{eq:14}
\end{eqnarray}
with 
\begin{eqnarray}
\mathcal{A}^{\textrm{Nonpeak}}&=&g_{\textrm{Nop}}\bar{u}(p_4)(\gamma\cdot \epsilon_{\psi} )v(p_5) \nonumber \\
&& \times (m_{p\bar{p}}-2m_p)^{a}(m_{B_s}-m_{J/\psi}-m_{p\bar{p}})^{b}, 
\end{eqnarray}
where the factor $(m_{p\bar{p}}-2m_p)^{a}(m_{B_s}-m_{J/\psi}-m_{p\bar{p}})^{b}$ \cite{BESIII:2013qmu} is introduced to phenomenologically absorb the nonpeaking contributions from other intermediate light mesonic states and  the index $i$ corresponds to the selected intermediate $f_2$ resonance and $\phi_i$ is the relative phase angle.

\begin{table}[b]
  	\caption{ The light isoscalar tensor meson family around 2.0 GeV. The listed masses and widths are in the unit of MeV.}
  	\setlength{\tabcolsep}{4.6mm}{
  	\begin{tabular}{cccccc}
			\toprule[1.2pt] 
		 $nL$	& $f_2$  meson &  Mass \cite{Ye:2012gu} & Width \cite{Ye:2012gu}  \\
			\toprule[0.8pt] 
        $3P$~($n\bar{n}$)     & $f_2(1910) $ & $1903 \pm 9$  & $196\pm31$  \\
      $1F$~($n\bar{n}$)      & $f_2(2010)$  & $2011^{+60}_{-80}$  & $202\pm60$  \\
      $3P$~($s\bar{s}$)      & $f_2(2150)$  & $2157\pm12$   & $152\pm30$  \\
      $4P$~($n\bar{n}$)     & $f_2(2200)$  & 2220  & 50 \\

\bottomrule[1.2pt]

\end{tabular}\label{table:f2}}
  \end{table}

With the above preparation, in the following, we analyze three invariant mass spectra of $B_s \to J/\psi p\bar{p}$ in two ways; one is the case in which we only include the reflections from highly exciting $f_2$ states, and another is the case in which we include the pentaquark contributions of $P_c(4380)^+$ and $P_c(4380)^-$ together with $f_2$ states, which we name as the scheme-I and scheme-II, respectively, and they are presented in Fig. \ref{fig:fit}.

In the scheme-I fit, it can be seen in Fig. \ref{fig:fit} that the reflections from the highly exciting $f_2$ states around 2.0 GeV plotted by the dotted line can really produce a peak line shape near 4.34 GeV in the $J/\psi p$ or $J/\psi \bar{p}$ mass spectrum.
However, because of the conservation of the total decay width of $B_s \to J/\psi p\bar{p}$, a strong enough reflection peak to match the experimental data of the  $P_c(4337)$  structure calls for an excessive distribution of the $f_2$ states in the $p\bar{p}$ mass spectrum as shown in Fig. \ref{fig:fit}. As a result, we conclude that this $J/\psi p$ structure near 4.34 GeV cannot be described by only considering the reflection contributions from light meson states, and accordingly an exotic pentaquark state that directly decays into final states of $J/\psi p$ is needed in our theoretical analysis.  In the following, we focus on studying whether the inclusion of $P_c(4380)^{+}$ and $P_c(4380)^{-}$ with the assumption of $J^{P}=3/2^{-}$ can improve the description to the line shape of the $p\bar{p}$, $J/\psi p$ and $J/\psi \bar{p}$ invariant mass spectra of $B_s \to J/\psi p\bar{p}$.

\begin{figure}[t]
	\includegraphics[width=7cm,keepaspectratio]{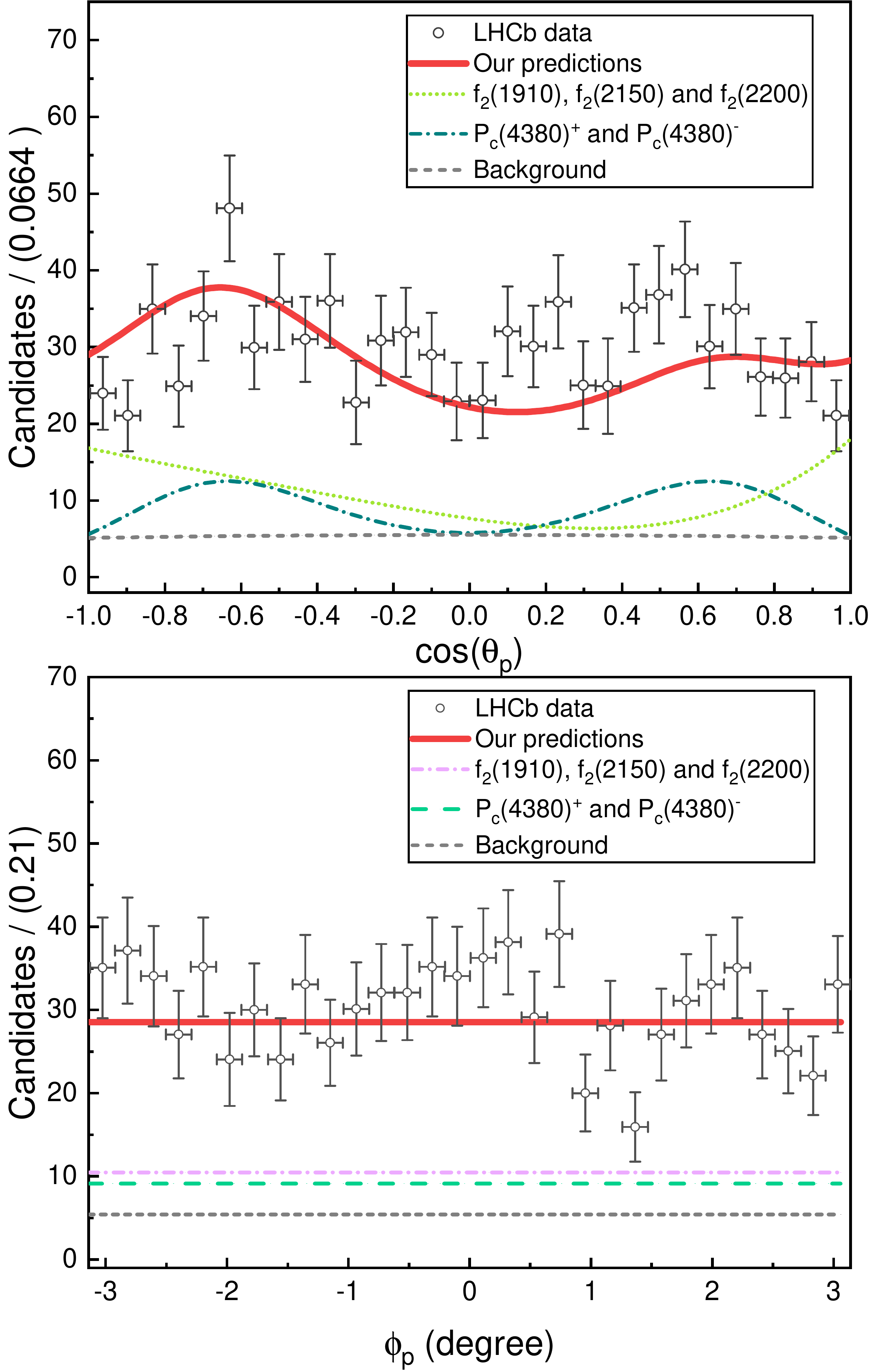}
	\caption{ The comparison between the predicted angular distributions  of decay process $B_s \to J/\psi p\bar{p}$  and the corresponding experimental data. \label{fig:angular} }
\end{figure}

In the scheme-II fit, we add the pentaquark amplitude term $e^{i\phi_4}(\mathcal{A}(B_s \to (P_c^{*+}(4380)\to J/\psi p)\bar{p})+\mathcal{A}(B_s \to (P_c^{*-}(4380)\to J/\psi \bar{p})p))$ into the total amplitude of Eq. (\ref{eq:14}),  and the corresponding combined analysis to the line shape of the invariant mass spectra of $B_s \to J/\psi p\bar{p}$ is shown in Fig. \ref{fig:fit}. The obtained $\chi^2/d.o.f.$ value is 1.5 and the corresponding parameters are listed in Table \ref{table:para}.  It is worth mentioning that the total amplitude is independent of the phase angle $\phi_1$ associated with $f_2(1910)$ due to the vanishing of  relevant interference term. Additionally, limited by the numerical method of the theoretical analysis, it is hard to fit the parameter $a$ and $b$ in the direct nonpeaking amplitude, and we test typical values of $a=0.457$ and $b=1.605$. In the new scheme, both two possible enhancements around 2.0 and 2.1 GeV and a dip near 2.2 GeV on the invariant mass spectrum of $p\bar{p}$ can be just described well by the $f_2(1910)$, $f_2(2150)$, and $f_2(2200)$ states successfully. At the same time, the intriguing $J/\psi p$ and $J/\psi \bar{p}$ structures around 4.34 GeV can be basically reproduced by the combined contributions from the pentaquark $P_c(4380)^{\pm}$ with $J^{P}=3/2^{-}$ and reflection peaks of light isoscalar tensor meson states. Consequently, our theoretical analysis of the invariant mass spectra of $B_s \to J/\psi p\bar{p}$ provides a direct evidence for the existence of $P_c(4380)^{\pm}$.

 We notice an interesting point that $P_c(4380)$ was generally predicted as a bound state of $S$-wave $\bar{D}\Sigma_c^{*}$ molecule in the past theoretical studies \cite{Chen:2015loa,Shen:2016tzq,He:2015cea,Lu:2016nnt,Lin:2017mtz,Lin:2019qiv,Du:2021fmf,Sakai:2019qph,Xiao:2020frg,Yalikun:2021bfm}, where the spin parity of $J^{P}=3/2^{-}$ can be naturally obtained. Thus, once the opinion is confirmed in the future precise measurement that the $J/\psi p$ structure around 4.34 GeV in  $B_s \to J/\psi p\bar{p}$ can be ascribed to the dominant $P_c(4380)^{\pm}$ contribution, then, this will be strong evidence to support the molecular nature of the $P_c(4380)$ state.  In order to further test our perspective, utilizing the fitting parameters listed in Table \ref{table:para}, we predict the angular distribution information of $B_s \to J/\psi p\bar{p}$, which can be directly compared with the relevant experimental data.

The predicted differential decay widths  of $B_s \to J/\psi p\bar{p}$ vs  $\textrm{cos}~\theta_{p}$ and $\phi_{p}$ are presented in Fig. \ref{fig:angular}. Here, $\theta_{p}$ and $\phi_{p}$ mean the solid angles between the proton direction in the rest frame of the $p\bar{p}$ system and a bachelor $J/\psi$ particle.
For the angular distribution against $\textrm{cos}~\theta_{p}$, it can be seen that the LHCb experimental data can well match our theoretical predictions. Interestingly, the line shape of the contributions of $P_c(4380)^{+}$ and $P_c(4380)^{-}$ is obviously different from those of the reflections from the $f_2$ states. Therefore, we suggest experimentalists to exactly measure the angular distribution of $B_s \to J/\psi p\bar{p}$ on $\textrm{cos}~\theta_{p}$, which should be helpful to distinguish the contribution  of $P_c(4380)^{\pm}$. The angular distribution of $B_s \to J/\psi p\bar{p}$ on $\phi_{p}$ should be symmetric for different types of production processes, so the theoretical prediction for the distribution line shape is a straight line, which is also consistent with the corresponding LHCb data within the margin of errors. Combined with the description of the invariant mass spectra and  angular distributions for $B_s \to J/\psi p\bar{p}$, we actually give a new perspective to decode the nature of the newly observed $J/\psi p$ structure near 4.34 GeV. Simultaneously, this points out possible new evidence for the existence of $P_c(4380)^{\pm}$ with $J^{P}=3/2^{-}$ from the measurements of $B_s \to J/\psi p\bar{p}$, and specifically this quantum number assignment agrees with the molecular explanation of $P_c(4380)$.

\begin{table}[t]
  	\caption{ The parameters of the fit to three invariant mass spectra of $B_s \to J/\psi p\bar{p}$  in the scheme-I and scheme-II, where $\phi_{2}$, $\phi_{3}$ and $\phi_{4}$ correspond to the phase angles associated with $f_2(2150)$, $f_2(2200)$, and $P_c(4380)$, respectively. Here, the $\chi^2/d.o.f.$ of the scheme-I and scheme-II fits are 4.3 and 1.5, respectively. }
  	\setlength{\tabcolsep}{5.0mm}{
  	\begin{tabular}{cccccccccc}
			\toprule[1.2pt] 
	Parameters	&  Scheme-I 	& Scheme-II  \\	
	\toprule[0.8pt] 
	$|g_{f_2(1910)}/g_{\textrm{Nop}}|$	&  $0.072 \pm 0.006 $ 	& $0.099\pm0.014$  \\	
	$|g_{f_2(2150)}/g_{\textrm{Nop}}|$	&  $0.108 \pm 0.005 $ 	& $0.142\pm0.015$  \\	
	$|g_{f_2(2200)}/g_{\textrm{Nop}}|$	&  $0.219 \pm 0.015$ 	& $0.371\pm0.049$  \\	
	$|g_{P_c(4380)}/g_{\textrm{Nop}}|$	&  $\cdots$ 	& $0.0013\pm0.0002$  \\	
	$\phi_{2}$	& $3.57\pm0.04$	& $3.71\pm0.04$  \\	
	$\phi_{3}$	&  $2.39\pm0.11$ 	& $2.04\pm0.09$  \\	
	$\phi_{4}$	&  $\cdots$ 	& $1.43\pm0.16$  \\	
	$a$~(Input)	&  0.457 	& 0.457  \\	
	$b$~(Input)	&  1.605 	& 1.605  \\	
			
\bottomrule[1.2pt]

\end{tabular}\label{table:para}}
  \end{table}

\section{Conclusions}\label{sec4}

Searching for the exotic multiquark hadronic states has been an extremely significant issue of hadron physics. In the past several decades, the observations of many charmoniumlike $XYZ$ states and three $P_c$ states in $\Lambda_b \to J/\psi p K$ have motivated extensive exploration of hidden-charm tetraquark and pentaquark hadrons \cite{Liu:2019zoy,Chen:2016qju}. An exciting advance in the field of multiquark states continues. Very recently, the LHCb collaboration released the  measurements of the decay process $B_s \to J/\psi p\bar{p}$ \cite{LHCb:2021chn}, where a new pentaquark candidate $P_c(4337)$ with the statistical significance of $3.1-3.7$ $\sigma$ was observed in the $J/\psi p$ invariant mass spectrum. Contrary to previously reported $P_c(4312)$, $P_c(4440)$, and $P_c(4457)$ in $\Lambda_b \to J/\psi p K$ \cite{LHCb:2019kea}, which can naturally correspond to the loosely bound $\bar{D}^{(*)}\Sigma_c$ \cite{Chen:2019asm,Liu:2019tjn,Yamaguchi:2019seo,Chen:2019bip,Xiao:2019aya,Meng:2019ilv,PavonValderrama:2019nbk,He:2019ify,Du:2019pij,Burns:2019iih,Wang:2019ato}, it is not an easy task to explain the nature of this newly observed $J/\psi p$ structure around 4.34 GeV \cite{Yan:2021nio,Liu:2021ixf,Liu:2021efc,Nakamura:2021dix}.

In this work, we have proposed an unconventional explanation to the observations of the  $B_s \to J/\psi p\bar{p}$, where two possibilities have been mainly investigated. The first possibility is that this $J/\psi p$ enhancement around 4.34 GeV can be caused by a nearby pentaquark $P_c(4380)$ state because this near-threshold resonance may deviate from the standard Breit-Wigner distribution. By a concrete research, we have found that the spin-parity assignment of $J^{P}=3/2^{-}$ to $P_c(4380)$ can indeed produce a resonance distribution in the $J/\psi p$ mass spectrum with a peak position near 4.34 GeV.  The second possibility is that  the reflection mechanism from the intermediate $f_2$ meson decays into $p\bar{p}$, which can mimic the resonance signal in the $J/\psi p$ distribution. After careful comparison between the LHCb data on the $p\bar{p}$ invariant mass spectrum of $B_s \to J/\psi p\bar{p}$, we have found that several suspected signals can be just related to three highly exciting $f_2(1910)$, $f_2(2150)$, and $f_2(2200)$ states around 2.0 GeV. Based on these research results, we have further carried out a combined fit to the LHCb data of three invariant mass spectra of $B_s \to J/\psi p\bar{p}$ with two fit scenarios. It can be concluded that the only reflection contributions cannot simultaneously describe the line shapes of three invariant mass spectra, and inclusion of the contributions of $P_c(4380)^{+}$ and $P_c(4380)^{-}$ with $J^{P}=3/2^{-}$ is absolutely necessary, which can largely improve whole fitting quality. Furthermore, our predictions for the angular distributions of $B_s \to J/\psi p\bar{p}$ can also match the corresponding LHCb data well. 

The theoretical analysis here means that our perspective of decoding the $P_c(4337)$ structure provides a new evidence to support the existence of the missing $P_c(4380)$ state in the 2019 updated data of $\Lambda_b \to J/\psi p K$ \cite{LHCb:2019kea}. More importantly, the assignment of $J^{P}=3/2^{-}$  is a natural result of a loosely $S$-wave $\bar{D}\Sigma_c^*$ molecular bound state. Thus, the determination of this quantum number agrees with the $\bar{D}\Sigma_c^*$ molecular configuration  of $P_c(4380)$. Together with three molecular pentaquark candidates $P_c(4312)$, $P_c(4440)$, and $P_c(4457)$, the $P_c(4380)$ state can further enlarge the present molecular pentaquark hadron family.  It is worth expecting that our argument can be tested in future precise measurements of $B_s \to J/\psi p\bar{p}$, especially on the run III data of LHC \cite{LHCb:2018roe}.


\medskip

\section*{ACKNOWLEDGMENTS}

This work is supported by the China National Funds for Distinguished Young Scientists under Grant No. 11825503, National Key Research and Development Program of China under Contract No. 2020YFA0406400, 111 Project under Grant No. B20063, and National Natural Science Foundation of China under Grant No. 12047501.


\begin{thebibliography}{199}


\bibitem{LHCb:2021chn}
R.~Aaij \textit{et al.} [LHCb Collaboration],
Evidence for a new structure in the $J/\psi p$ and $J/\psi \bar{p}$ systems in $B_s^0 \to J/\psi p \bar{p}$ decays,
arXiv:2108.04720.


\bibitem{LHCb:2019kea}
R.~Aaij \textit{et al.} [LHCb Collaboration],
Observation of a narrow pentaquark state, $P_c(4312)^+$, and of two-peak structure of the $P_c(4450)^+$,
Phys. Rev. Lett. \textbf{122}, 222001 (2019).


\bibitem{Wu:2010jy}
J.~J.~Wu, R.~Molina, E.~Oset and B.~S.~Zou,
Prediction of narrow $N^*$ and $\Lambda^*$ resonances with hidden charm above 4 GeV,
Phys. Rev. Lett. \textbf{105}, 232001 (2010).


\bibitem{Wang:2011rga}
W.~L.~Wang, F.~Huang, Z.~Y.~Zhang and B.~S.~Zou,
$\Sigma_c \bar{D}$ and $\Lambda_c \bar{D}$ states in a chiral quark model,
Phys. Rev. C \textbf{84}, 015203 (2011).


\bibitem{Yang:2011wz}
Z.~C.~Yang, Z.~F.~Sun, J.~He, X.~Liu and S.~L.~Zhu,
The possible hidden-charm molecular baryons composed of anti-charmed meson and charmed baryon,
Chin. Phys. C \textbf{36}, 6 (2012).


\bibitem{Uchino:2015uha}
T.~Uchino, W.~H.~Liang and E.~Oset,
Baryon states with hidden charm in the extended local hidden gauge approach,
Eur. Phys. J. A \textbf{52}, 43 (2016).


\bibitem{Karliner:2015ina}
M.~Karliner and J.~L.~Rosner,
New Exotic Meson and Baryon Resonances from Doubly-Heavy Hadronic Molecules,
Phys. Rev. Lett. \textbf{115},  122001 (2015).


\bibitem{Chen:2019asm}
R.~Chen, Z.~F.~Sun, X.~Liu and S.~L.~Zhu,
Strong LHCb evidence supporting the existence of the hidden-charm molecular pentaquarks,
Phys. Rev. D \textbf{100},  011502 (2019).



\bibitem{Liu:2019tjn}
M.~Z.~Liu, Y.~W.~Pan, F.~Z.~Peng, M.~S\'anchez S\'anchez, L.~S.~Geng, A.~Hosaka and M.~Pavon Valderrama,
Emergence of a complete heavy-quark spin symmetry multiplet: seven molecular pentaquarks in light of the latest LHCb analysis,
Phys. Rev. Lett. \textbf{122},  242001 (2019).


\bibitem{Yamaguchi:2019seo}
Y.~Yamaguchi, H.~Garc\'\i{}a-Tecocoatzi, A.~Giachino, A.~Hosaka, E.~Santopinto, S.~Takeuchi and M.~Takizawa,
$P_c$ pentaquarks with chiral tensor and quark dynamics,
Phys. Rev. D \textbf{101},  091502 (2020).


\bibitem{Chen:2019bip}
H.~X.~Chen, W.~Chen and S.~L.~Zhu,
Possible interpretations of the $P_c(4312)$, $P_c(4440)$, and $P_c(4457)$,
Phys. Rev. D \textbf{100},  051501 (2019).


\bibitem{Xiao:2019aya}
C.~W.~Xiao, J.~Nieves and E.~Oset,
Heavy quark spin symmetric molecular states from ${\bar D}^{(*)}\Sigma_c^{(*)}$ and other coupled channels in the light of the recent LHCb pentaquarks,
Phys. Rev. D \textbf{100},  014021 (2019).


\bibitem{Meng:2019ilv}
L.~Meng, B.~Wang, G.~J.~Wang and S.~L.~Zhu,
The hidden charm pentaquark states and $\Sigma_c\bar{D}^{(*)}$ interaction in chiral perturbation theory,
Phys. Rev. D \textbf{100},  014031 (2019).


\bibitem{PavonValderrama:2019nbk}
M.~Pavon Valderrama,
One pion exchange and the quantum numbers of the P$_c$(4440) and P$_c$(4457) pentaquarks,
Phys. Rev. D \textbf{100},  094028 (2019).


\bibitem{He:2019ify}
J.~He,
Study of $P_c(4457)$, $P_c(4440)$, and $P_c(4312)$ in a quasipotential Bethe-Salpeter equation approach,
Eur. Phys. J. C \textbf{79},  393 (2019).


\bibitem{Du:2019pij}
M.~L.~Du, V.~Baru, F.~K.~Guo, C.~Hanhart, U.~G.~Mei\ss{}ner, J.~A.~Oller and Q.~Wang,
Interpretation of the LHCb $P_c$ States as Hadronic Molecules and Hints of a Narrow $P_c(4380)$,
Phys. Rev. Lett. \textbf{124},  072001 (2020).


\bibitem{Burns:2019iih}
T.~J.~Burns and E.~S.~Swanson,
Molecular interpretation of the $P_c$(4440) and $P_c$(4457) states,
Phys. Rev. D \textbf{100},  114033 (2019).


\bibitem{Wang:2019ato}
B.~Wang, L.~Meng and S.~L.~Zhu,
Hidden-charm and hidden-bottom molecular pentaquarks in chiral effective field theory,
J. High Energy Phys. \textbf{11} (2019) 108.



\bibitem{Yan:2021nio}
M.~J.~Yan, F.~Z.~Peng, M.~S.~S\'anchez and M.~P.~Valderrama,
Interpretations of the new LHCb $P_c(4337)^+$ pentaquark state,
arXiv:2108.05306.


\bibitem{Liu:2021ixf}
Y.~Liu, M.~A.~Nowak and I.~Zahed,
Holographic charm and bottom pentaquarks II: Open and hidden decay widths,
arXiv:2108.07074 [Phys. Rev. D (to be published)].


\bibitem{Liu:2021efc}
Y.~Liu, K.~A.~Mamo, M.~A.~Nowak and I.~Zahed,
Holographic charm and bottom pentaquarks III: Excitations through photo-production of heavy mesons,
arXiv:2109.03103 [Phys. Rev. D (to be published)].


\bibitem{Nakamura:2021dix}
S.~X.~Nakamura, A.~Hosaka and Y.~Yamaguchi,
$P_c(4312)^+$ and $P_c(4337)^+$ as interfering $\Sigma_c\bar{D}$ and $\Lambda_c\bar{D}^{*}$ (anomalous) threshold cusps,
arXiv:2109.15235.

\bibitem{Giron}
J.~F.~Giron and R.~F.~Lebed, 
Fine Structure of Pentaquark Multiplets in the Dynamical Diquark Model,
arXiv:2110.05557.



\bibitem{LHCb:2015yax}
R.~Aaij \textit{et al.} [LHCb Collaboration],
Observation of $J/\psi p$ Resonances Consistent with Pentaquark States in $\Lambda_b^0 \to J/\psi K^- p$ Decays,
Phys. Rev. Lett. \textbf{115}, 072001 (2015).



\bibitem{Wang:2020axi}
J.~Z.~Wang, D.~Y.~Chen, X.~Liu and T.~Matsuki,
Universal non-resonant explanation to charmoniumlike structures $Z_c(3885)$ and $Z_c(4025)$,
Eur. Phys. J. C \textbf{80}, 1040 (2020).


\bibitem{Wang:2020dmv}
J.~Z.~Wang, D.~Y.~Chen, X.~Liu and T.~Matsuki,
Mapping a new cluster of charmoniumlike structures at $e^+ e^-$ collisions,
Phys. Lett. B \textbf{817}, 136345 (2021).


\bibitem{Wang:2020kej}
J.~Z.~Wang, Q.~S.~Zhou, X.~Liu and T.~Matsuki,
Toward charged $Z_{cs}(3985)$ structure under a reflection mechanism,
Eur. Phys. J. C \textbf{81},  51 (2021).


\bibitem{Belle:2003nnu}
S.~K.~Choi \textit{et al.} [Belle Collaboration],
Observation of a narrow charmonium-like state in exclusive $B^\pm \to K^\pm \pi^+ \pi^- J/\psi$ decays,
Phys. Rev. Lett. \textbf{91}, 262001 (2003).



\bibitem{Belle:2004lle}
K.~Abe \textit{et al.} [Belle Collaboration],
Observation of a near-threshold $\omega J/\psi$ mass enhancement in exclusive $B \to K \omega J/\psi$ decays,
Phys. Rev. Lett. \textbf{94}, 182002 (2005).


\bibitem{Belle:2007hrb}
S.~K.~Choi \textit{et al.} [Belle Collaboration],
Observation of a resonance-like structure in the $\pi^\pm \psi^\prime$ mass distribution in exclusive $B \to K \pi^\pm \psi^\prime$ decays,
Phys. Rev. Lett. \textbf{100}, 142001 (2008).


\bibitem{CMS:2013jru}
S.~Chatrchyan \textit{et al.} [CMS Collaboration],
Observation of a Peaking Structure in the $J/\psi \phi$ Mass Spectrum from $B^{\pm} \to J/\psi \phi K^{\pm}$ Decays,
Phys. Lett. B \textbf{734}, 261 (2014).


\bibitem{LHCb:2021uow}
R.~Aaij \textit{et al.} [LHCb Collaboration],
Observation of New Resonances Decaying to $J/\psi K^+$+ and $J/\psi \phi$,
Phys. Rev. Lett. \textbf{127},  082001 (2021).


\bibitem{Wang:2021aql}
F.~L.~Wang, X.~D.~Yang, R.~Chen and X.~Liu,
Correlation of the hidden-charm molecular tetraquarks and the charmonium-like structures existing in the $B\to XYZ+K$,
Phys. Rev. D \textbf{104}, 094010 (2021).



\bibitem{Liu:2019zoy}
  Y.~R.~Liu, H.~X.~Chen, W.~Chen, X.~Liu and S.~L.~Zhu,
  Pentaquark and Tetraquark states,
  Prog.\ Part.\ Nucl.\ Phys.\  {\bf 107}, 237 (2019).



\bibitem{Chen:2016qju}
  H.~X.~Chen, W.~Chen, X.~Liu and S.~L.~Zhu,
  The hidden-charm pentaquark and tetraquark states,
  Phys.\ Rep.\  {\bf 639}, 1 (2016).
  
  
  
\bibitem{ParticleDataGroup:2020ssz}
P.~A.~Zyla \textit{et al.} [Particle Data Group],
Review of Particle Physics,
Prog. Theor. Exp. Phys. \textbf{2020},  083C01 (2020).


\bibitem{Belle:2011phz}
J.~Li \textit{et al.} [Belle Collaboration],
Observation of $B_s^0\to J/\psi f_0(980)$ and Evidence for $B_s^0\to J/\psi f_0(1370)$,
Phys. Rev. Lett. \textbf{106}, 121802 (2011).


\bibitem{LHCb:2014ooi}
R.~Aaij \textit{et al.} [LHCb Collaboration],
Measurement of resonant and CP components in $\bar{B}_s^0\to J/\psi\pi^+\pi^-$ decays,
Phys. Rev. D \textbf{89},  092006 (2014).


\bibitem{LHCb:2012ae}
R.~Aaij \textit{et al.} [LHCb Collaboration],
Analysis of the resonant components in $B_s \to J/\psi\pi^+\pi^-$,
Phys. Rev. D \textbf{86}, 052006 (2012).


\bibitem{Tsushima:1996xc}
K.~Tsushima, A.~Sibirtsev and A.~W.~Thomas,
Resonance model study of strangeness production in $pp$ collisions,
Phys. Lett. B \textbf{390}, 29 (1997).


\bibitem{Tsushima:1998jz}
K.~Tsushima, A.~Sibirtsev, A.~W.~Thomas and G.~Q.~Li,
Resonance model study of kaon production in baryon baryon reactions for heavy ion collisions,
Phys. Rev. C \textbf{59}, 369 (1999)
[erratum: Phys. Rev. C \textbf{61}, 029903 (2000)].


\bibitem{Zou:2002yy}
B.~S.~Zou and F.~Hussain,
Covariant $L-S$ scheme for the effective $N^*NM$ couplings,
Phys. Rev. C \textbf{67}, 015204 (2003).


\bibitem{Wu:2009md}
J.~J.~Wu, Z.~Ouyang and B.~S.~Zou,
Proposal for Studying $N^*$ Resonances with $\bar{p} p \to  \bar{p} n \pi^+$ Reaction,
Phys. Rev. C \textbf{80}, 045211 (2009).


\bibitem{Wang:2017sxq}
J.~Z.~Wang, H.~Xu, J.~J.~Xie and X.~Liu,
Production of the charmoniumlike state Y(4220) through the $p\bar{p} \to Y(4220) \pi^0$ reaction,
Phys. Rev. D \textbf{96}, 094004 (2017).


\bibitem{Colangelo:2010te}
P.~Colangelo and F.~De Fazio,
Open charm meson spectroscopy: Where to place the latest piece of the puzzle,
Phys. Rev. D \textbf{81}, 094001 (2010).


\bibitem{Yan:1999fn}
Y.~Yan and R.~Tegen,
Baryon exchange and meson pole diagrams in $p \bar{p} \to \bar{K} K$, $ \pi^+ \pi^-$,
Nucl. Phys. A\textbf{648}, 89 (1999).


\bibitem{Goldberg:1968zza}
H.~Goldberg,
Backward ($\theta=180^{\circ}$) $\pi N$ Dispersion Relations: Applications to the Interference Model, $P$ and $P^{\prime}$ Trajectories, and to the Mechanical Form Factors of the Nucleon,
Phys. Rev. \textbf{171}, 1485 (1968).


\bibitem{Feuster:1997pq}
T.~Feuster and U.~Mosel,
A Unitary model for meson nucleon scattering,
Phys. Rev. C \textbf{58}, 457 (1998).


\bibitem{Haberzettl:1998aqi}
H.~Haberzettl, C.~Bennhold, T.~Mart and T.~Feuster,
Gauge-invariant tree-level photoproduction amplitudes with form factors,
Phys. Rev. C \textbf{58},  R40 (1998).


\bibitem{Yoshimoto:1999dr}
T.~Yoshimoto, T.~Sato, M.~Arima and T.~S.~H.~Lee,
Dynamical test of constituent quark models with $\pi N$ reactions,
Phys. Rev. C \textbf{61}, 065203 (2000).


\bibitem{Oh:2000zi}
Y.~s.~Oh, A.~I.~Titov and T.~S.~H.~Lee,
Nucleon resonances in omega photoproduction,
Phys. Rev. C \textbf{63}, 025201 (2001).


\bibitem{Ye:2012gu}
Z.~C.~Ye, X.~Wang, X.~Liu and Q.~Zhao,
The mass spectrum and strong decays of isoscalar tensor mesons,
Phys. Rev. D \textbf{86}, 054025 (2012).


\bibitem{CrystalBarrel:2006mhj}
C.~Amsler \textit{et al.} [Crystal Barrel Collaboration],
Study of $K\bar{K}$ resonances in $\bar{p} p \to K^+ K^- \pi^0$ at 900 MeV/$c$ and 1640 MeV/$c$,
Phys. Lett. B \textbf{639}, 165 (2006).


\bibitem{Uman:2006xb}
I.~Uman, D.~Joffe, Z.~Metreveli, K.~K.~Seth, A.~Tomaradze and P.~K.~Zweber,
Light quark resonances in $p\bar{p}$ annihilations at 5.2 GeV/$c$,
Phys. Rev. D \textbf{73}, 052009 (2006).


\bibitem{CrystalBarrel:1999hmx}
A.~Abele \textit{et al.} [Crystal Barrel Collaboration],
Observation of resonances in the reaction $\bar{p} p \to \pi^0 \eta \eta$ at 1.94 GeV/$c$,
Eur. Phys. J. C \textbf{8}, 67 (1999).


\bibitem{Dulude:1978kt}
R.~S.~Dulude, R.~E.~Lanou, J.~T.~Massimo, D.~C.~Peaslee, R.~K.~Thornton, D.~S.~Barton, M.~Marx, B.~A.~Nelson, L.~Rosenson and C.~DeMarzo, \textit{et al.}
Observation of Structure in $\bar{p} p \to \pi^0 \pi^0$,
Phys. Lett. \textbf{79}B, 335 (1978).


\bibitem{Evangelista:1997be}
C.~Evangelista, A.~Palano, D.~Drijard, N.~H.~Hamann, R.~T.~Jones, B.~Mouellic, S.~Ohlsson, J.~M.~Perreau, W.~Eyrich and M.~Moosburger, \textit{et al.}
Measurement of the $\bar{p} p \to K_{s} K_{s}$ reaction from 0.6 GeV/$c$ to 1.9 GeV/$c$,
Phys. Rev. D \textbf{56}, 3803 (1997).


\bibitem{BESIII:2013qmu}
M.~Ablikim \textit{et al.} [BESIII Collaboration],
Observation of a charged $(D\bar{D}^{*})^\pm$ mass peak in $e^{+}e^{-} \to \pi D\bar{D}^{*}$ at $\sqrt{s} =$ 4.26 GeV,
Phys. Rev. Lett. \textbf{112},  022001 (2014).


\bibitem{Chen:2015loa}
R.~Chen, X.~Liu, X.~Q.~Li and S.~L.~Zhu,
Identifying exotic hidden-charm pentaquarks,
Phys. Rev. Lett. \textbf{115}, 132002 (2015).


\bibitem{Shen:2016tzq}
C.~W.~Shen, F.~K.~Guo, J.~J.~Xie and B.~S.~Zou,
Disentangling the hadronic molecule nature of the $P_c(4380)$ pentaquark-like structure,
Nucl. Phys. A\textbf{954}, 393 (2016).


\bibitem{He:2015cea}
J.~He,
$\bar{D}\Sigma^*_c$ and $\bar{D}^*\Sigma_c$ interactions and the LHCb hidden-charmed pentaquarks,
Phys. Lett. B \textbf{753}, 547 (2016).


\bibitem{Lu:2016nnt}
Q.~F.~L\"u and Y.~B.~Dong,
Strong decay mode $J/\psi p$ of hidden charm pentaquark states $P_c^+(4380)$ and $P_c^+(4450)$ in $\Sigma_c \bar{D}^*$ molecular scenario,
Phys. Rev. D \textbf{93}, 074020 (2016).


\bibitem{Lin:2017mtz}
Y.~H.~Lin, C.~W.~Shen, F.~K.~Guo and B.~S.~Zou,
Decay behaviors of the $P_c$ hadronic molecules,
Phys. Rev. D \textbf{95}, 114017 (2017).


\bibitem{Lin:2019qiv}
Y.~H.~Lin and B.~S.~Zou,
Strong decays of the latest LHCb pentaquark candidates in hadronic molecule pictures,
Phys. Rev. D \textbf{100}, 056005 (2019).


\bibitem{Du:2021fmf}
M.~L.~Du, V.~Baru, F.~K.~Guo, C.~Hanhart, U.~G.~Mei\ss{}ner, J.~A.~Oller and Q.~Wang,
Revisiting the nature of the P$_{c}$ pentaquarks,
J. High Energy Phys. \textbf{08} (2021) 157.


\bibitem{Sakai:2019qph}
S.~Sakai, H.~J.~Jing and F.~K.~Guo,
Decays of $P_c$ into $J/\psi N$ and $\eta_cN$ with heavy quark spin symmetry,
Phys. Rev. D \textbf{100},  074007 (2019).


\bibitem{Xiao:2020frg}
C.~W.~Xiao, J.~X.~Lu, J.~J.~Wu and L.~S.~Geng,
How to reveal the nature of three or more pentaquark states,
Phys. Rev. D \textbf{102}, 056018 (2020).


\bibitem{Yalikun:2021bfm}
N.~Yalikun, Y.~H.~Lin, F.~K.~Guo, Y.~Kamiya and B.~S.~Zou,
Coupled Channel Effects of the $\Sigma_c^{(*)}\bar{D}^{(*)}$-$\Lambda_c(2595)\bar{D}$ System and Molecular Nature of the $P_c$ Pentaquark States from One-Boson Exchange Model,
arXiv:2109.03504.


\bibitem{LHCb:2018roe}
R.~Aaij \textit{et al.} [LHCb Collaboration],
Physics case for an LHCb Upgrade II - Opportunities in flavour physics, and beyond, in the HL-LHC era,
arXiv:1808.08865.




\end{thebibliography}
\end{document}